\documentclass[aps,pra,reprint,onecolumn]{revtex4-2}

\usepackage[utf8]{inputenc}
\usepackage{amsthm}
\usepackage{amsmath}
\usepackage{color}
\usepackage{hyperref}
\usepackage{graphicx,xcolor}
\usepackage{amssymb}
\usepackage{bbold}

\newcommand{\figheight}{4.9cm}

\newcommand{\ii}{\mathrm{i}} 
\newcommand{\eul}{\mathrm{e}} 
\newcommand{\diff}{\mathrm{d}} 
\newcommand{\id}{\mathbb{1}} 
\newcommand{\vz}{\boldsymbol{z}} 
\newcommand{\vac}{\boldsymbol{0}} 

\newcommand{\expval}[1]{{\mathcal{M}}\left[#1\right]} 
\newcommand{\expvalbeta}[1]{\mathcal{M}_{\beta}\left[#1\right]} 
\newcommand{\bvec}[1]{\boldsymbol{#1}} 

\newcommand{\ket}[1]{|#1\rangle} 
\newcommand{\bra}[1]{\langle#1|} 
\newcommand{\braket}[1]{\langle#1\rangle} 
\newcommand{\ketbra}[2]{|#1\rangle\!\langle #2|} 
\newcommand{\tr}{\mathrm{tr}} 

\newcommand{\emphquote}[1]{‘#1’} 
\begin{document}

\title{Non-Markovian Quantum State Diffusion for Spin Environments}
\author{Valentin Link}
\affiliation{Institut f{\"u}r Theoretische Physik, Technische Universit{\"a}t Dresden, 
D-01062, Dresden, Germany}
\date{\today}

\author{Kimmo Luoma}
\affiliation{Laboratory of Quantum Optics, Department of Physics and Astronomy, 
  University of Turku, FI-20014, Turun yliopisto, Finland}
\email{ktluom@utu.fi}

\author{Walter T Strunz}
\affiliation{Institut f{\"u}r Theoretische Physik, Technische Universit{\"a}t Dresden, 
D-01062, Dresden, Germany}
\email{walter.strunz@tu-dresden.de}

\begin{abstract}
We introduce an exact open system method to describe the dynamics of quantum systems that are strongly coupled to specific types of environments comprising of spins, such as central spin systems. Our theory is similar to the established non-Markovian quantum state diffusion (NMQSD) theory, but for a spin bath instead of a Gaussian bath.
The method allows us to represent the time-evolved reduced state of the system as an ensemble average of stochastically evolving pure states. We present a comprehensive theory for arbitrary linear spin environments at both zero and finite temperatures. Furthermore, we introduce a hierarchical expansion method that enables the numerical computation of the time evolution of the stochastic pure states, facilitating a numerical solution of the open system problem in relevant strong coupling regimes.
\end{abstract}
\maketitle

\section{Introduction}
Open quantum system theory is concerned with investigating the dynamics of a quantum system in contact with a large environment. This presents a challenging computational problem that requires the use of suitable approximations and efficient numerical methods. As research in this field advances, there has been a steady increase in the ability to address complex problems where the system is strongly coupled to structured environments \cite{deVega2017Jan}. However, as the strength of the system-environment interaction increases, the validity of the standard assumptions that underlie most open system methods may become questionable. Conversely, relaxing these assumptions may open up new possibilities for using open system theory to address problems that were previously only accessible through many-body methods \cite{Bortz2010Oct,Villazon2020Sep,Schliemann2003Dec,Dukelsky2004Aug}.

A crucial requirement for the majority of open quantum system methods is that the response of the environment to the system is Gaussian. This is a well-founded assumption because most commonly studied quantum environments are either linearly coupled Gaussian bosonic systems, as is typical for instance in quantum optics \cite{wallsBook}, or they consist of a large number of weakly coupled degrees of freedom so that basic central limit theorem arguments apply, yielding an effective Gaussian response \cite{makri99,Fernandez-Acebal2018Mar}. However, a Gaussian description may loose validity for a variety of strongly coupled many-body systems, even if their structure suggests that they could be treated as open systems \cite{Bramberger2020Jan}. One such example are central spin models, which describe the interaction between a central spin and an ensemble of independent environmental spins. These models have been extensively studied within many-body theory using both analytical and numerical methods, and have applications in nuclear magnetic resonance spectroscopy, sensing, and solid-state quantum information platforms \cite{Villazon2021Feb,Taylor2003May,Fowler-Wright2023Mar}. Central spin problems have a suitable structure for open system theory, with the environmental spins being identified as a bath for the central spin. However, standard open system methods cannot be applied to these systems because a bath consisting of spins does not follow Gaussian statistics. This holds for almost all established methods such as the hierarchical equations of motion (HEOM) \cite{Tanimura2020Jul}, the quasi-adiabatic path integral (QUAPI/TEMPO) \cite{Makri1995Mar,Strathearn2018Aug}, or time-evolving density matrices using orthogonal polynomials (TEDOPA) \cite{Prior2010Jul}.
Only recently, some new approaches have been proposed that enable a treatment of non-Gaussian baths in an open system framework, an extension of HEOM \cite{Hsieh2018Jan1,Hsieh2018Jan} and a process tensor method based on a matrix product state representation \cite{,Cygorek2021Jan}. 

It is the goal of this work to generalize another open system theory, the non-Markovian quantum state diffusion (NMQSD) \cite{Strunz1996}, in order to describe open system dynamics with spin baths. In contrast to other approaches that work with mixed states, NMQSD allows for a computation of open system dynamics via stochastic sampling of pure state trajectories. In the case of Gaussian baths, quantum state diffusion is known to have certain technical and analytical advantages over mixed state descriptions. In particular, it is based on a propagation of state vectors rather than operators, which can be a significant advantage in certain systems \cite{Flannigan2021Aug}. The method has been utilized successfully both for simulation of non-Markovian dynamics~\cite{suess14,Hartmann2017,Zhang2016Nov,Chen2022Mar}, and as a versatile tool to derive non-Markovian master equations or to treat integrable problems \cite{Strunz2001Jun,strunz04}. 
In this paper we derive a nontrivial generalization of this framework for open system problems with spin baths, such as central spin problems, at zero and finite temperature. We derive exact evolution equations for the NMQSD trajectories by utilizing a projection formalism that was introduced in Ref.~\cite{Link17}. Further, we show that it is possible to numerically compute the solution of this equation with a hierarchy method, thus allowing us to simulate the reduced system dynamics in spin bath problems via Monte-Carlo sampling of stochastic pure state trajectories.

The remainder of this work is structured as follows. We first introduce the open system model in Sec.~\ref{sec:model} and discuss differences to the well known Gaussian bath model. In the technical part of this work, we first give a brief review on spin coherent states in Sec.~\ref{sec:spin_coherent_states}, as these will play an essential role in the following. Readers familiar with spin coherent states may directly switch to Sec.~\ref{sec:stochastic_theory}, where we develop the extension of NMQSD theory, the main result of this work. As a demonstration we present an application to the solvable pure dephasing model in Sec.~\ref{sec:pure_dephasing}. In Sec.~\ref{sec:hierarchy_expansion} we introduce a method which allows us to numerically solve the dynamics of individual NMQSD trajectories and thus, in principle, to compute the reduced dynamics of open systems with spin baths. Finally, we present our conclusions in Sec.~\ref{sec:conclusions}. Various longer derivations are provided in the appendix.

\section{Spin Bath Model}\label{sec:model}
The spin bath model we like to consider consists of an arbitrary quantum system $S$ interacting linearly with an ensemble of $N$ independent bath spins $\vec{J}_\lambda$. The Hamiltonian reads explicitly
\begin{equation}\label{eq:spinbathH}
    H=H_S+\sum_{\lambda=1}^N g_\lambda \vec{L}\cdot \vec{J}_\lambda +\sum_{\lambda=1}^N \omega_\lambda(j-J^z_\lambda),
\end{equation}
where $H_S$ is the system Hamiltonian and $\vec{L}$ is a vector of hermitian operators (coupling operators) in the system Hilbert space. The dot product denotes the standard euclidian scalar product. Further, we assume that the spins have a fixed length $j=\frac{n}{2}$ where $n\in \mathbb{N}$ (note that $H$ preserves $J^2_\lambda$). The spin operators $J_\lambda^{a}$ satisfy the usual algebra
\begin{equation}\label{eq:angular_momentum_algebra}
    [J_\lambda^a,J_\lambda^b]=\ii\varepsilon^{abc}J_\lambda^c\,
\end{equation}
and spin operators acting on different spins commute. Throughout this paper, roman indices indicating the spin directions $x,y,z$ are written as upper indices for notational convenience, and summation is implied when such an index appears twice. For fixed length spins an additional relation $\vec{J}_\lambda\cdot\vec{J}_\lambda=j(j+1)$ holds, and the Hilbert space of a single spin is spanned by the $2j+1$ eigenstates of $J^z$ 
\begin{equation}
    J_\lambda^z\ket{j,m}_\lambda=m\ket{j,m}_\lambda.
\end{equation} 
Hamiltonian \eqref{eq:spinbathH} includes all common spin bath models, for instance condensed phase environments or decoherence models for quantum computing \cite{Hsieh2018Jan,Schliemann2003Dec,Villazon2021Feb,Taylor2003May}. Variants of model \eqref{eq:spinbathH} have been studied previously with different methods such as the multi configurational time-dependent Hartree-Fock method~\cite{Wang2012Dec,Gelman2004Jul}, a dissipation formalism~\cite{Zhang2015Jan}, analytically \cite{Dukelsky2004Aug,Schliemann2003Dec} or perturbatively~\cite{Lu2009Oct}. 

In the following we always consider an interaction picture with respect to the free Hamiltonian of the bath spins, which results in the following Hamiltonian
\begin{equation}\label{eq:hamiltonian}
    H(t)=H_S+\sum_{\lambda=1}^N g_\lambda \vec{L}\cdot O_\lambda(t)\vec{J}_\lambda.
\end{equation}
In this expression, the orthogonal matrices $O_\lambda(t)\in \mathrm{SO}(3)$ describe the bare rotation of the spins around the $z$-axis
\begin{equation}
   O_\lambda(t)=\begin{pmatrix}
            \cos\omega_\lambda t &\sin\omega_\lambda t&0\\
            -\sin\omega_\lambda t &\cos\omega_\lambda t & 0\\
            0 & 0 & 1
        \end{pmatrix}.
\end{equation}
Assuming a pure initial state of the system and a zero temperature spin bath, the full system and bath state $\ket{\Psi(t)}$ obeys unitary evolution from the Schrödinger equation
\begin{equation}\label{eq:SEq}
    \partial_t\ket{\Psi(t)}=-\ii H(t)\ket{\Psi(t)}.
\end{equation}
Explicitly, the zero temperature bath state is simply the state where all spins are pointing up $\ket{\boldsymbol{0}}\equiv\ket{j,j}^{\otimes N}$, and the full initial condition is $\ket{\Psi(0)}=\ket{\psi_0}\ket{\boldsymbol{0}}$ with an arbitrary pure system state $\ket{\psi_0}$. We show later that the finite temperature problem can be reduced to zero temperature by introducing additional stochastic thermal fluctuations to the Hamiltonian. Solving the Schrödinger equation \eqref{eq:SEq} is hard, because the size of the system and bath Hilbert space grows exponentially with the number of bath spins. Open system methods aim to resolve this issue by finding a description of the reduced dynamics within the system Hilbert space, taking the form of a closed theory for the reduced system state
\begin{equation}
    \rho_S(t)=\tr_E\ketbra{\Psi(t)}{\Psi(t)}.
\end{equation}

One can compare the spin bath model \eqref{eq:hamiltonian} to the Gaussian bath model that is usually considered in open quantum systems. It is well known that any environment with a Gaussian influence functional can be represented in terms of independent bosonic modes, so that the general open system Hamiltonian for a stationary Gaussian bath can be written as \cite{Yang2016Nov}
\begin{equation}\label{eq:H_Gaussian}
    H=H_S+\int \diff \omega \sum_{i,k} L_i ({g_{ik}(\omega)} \eul^{-\ii\omega t}b_k(\omega)+\mathrm{h.c.}),
\end{equation}
where $[ b_k(\omega),b_l^\dagger(\omega')]=\delta_{kl}\delta(\omega-\omega')$. Crucially, the bath coupling operators 
\begin{equation}
    B_i(t)=\int\diff\omega\sum_k({g_{ik}(\omega)} \eul^{-\ii\omega t}b_k(\omega)+\mathrm{h.c.})
\end{equation}
obey Gaussian statistics in the sense that all higher moments can be obtained via Wick's theorem from the second moment
\begin{equation}
     \braket{\mathrm{vac}|B_i(t)B_j(s)|\mathrm{vac}}=\alpha_{ij}(t-s),
\end{equation}
where $\alpha_{ij}(t)$ is the bath correlation function and $\ket{\mathrm{vac}}$ is the bare vacuum state of the bath. The spin bath model cannot be mapped to a Gaussian bath model because higher order cumulants of the bath coupling operators
\begin{equation}
    \vec{B}(t)=\sum_{\lambda=1}^N g_\lambda O_\lambda(t)\vec{J}_\lambda
\end{equation}
do not vanish (this is a direct consequence of the nontrival algebra \eqref{eq:angular_momentum_algebra}). For a strong system and bath coupling it may not be sufficient to treat the spin bath within a Gaussian approximation in order to accurately describe the system dynamics, see for instance the analysis in Ref.~\cite{Bramberger2020Jan}.

Non-Markovian quantum state diffusion (NMQSD) is an established theoretical framework to describe dynamics due to Gaussian environments, i.e.~Hamiltonian \eqref{eq:H_Gaussian}. It is a generalization of trajectory methods, such as quantum state diffusion or quantum jumps, to non-Markovian systems \cite{Strunz99}. NMQSD provides an unraveling of the system dynamics in terms of a stochastic process in the system Hilbert space $\ket{\psi(t)}$, such that the average reproduces the mixed system state
\begin{equation}\label{eq:NMQSD_avg}
    \rho_S(t)=\expval{ \ketbra{\psi(t)}{\psi(t)}}.
\end{equation}
The expectation value $\expval{...}$ above is to be understood with respect to all stochastic realizations of the process $\ket{\psi(t)}$, and can in practice be evaluated using Monte Carlo sampling of trajectories. 
This article presents a generalization of the NMQSD method to address spin bath problems described by the equation \eqref{eq:hamiltonian}. Unlike most established open systems methods, this new approach can handle spin baths in an exact manner. A detailed derivation of the theory is provided below. The main results include an exact unraveling of the reduced system state in terms of normalized pure states (Eq.~\eqref{eq:SCunrav_nonlinear}) obeying a closed evolution equation (Eq.~\eqref{eq:Feshbach}).

\section{Properties of Spin Coherent States}\label{sec:spin_coherent_states}

The microscopic derivation of NMQSD relies on coherent states, which are also a crucial element in the generalization to spin baths that we aim to establish. Therefore, it is important to briefly introduce the essential properties of spin coherent states. Spin coherent states can be introduced in various ways, depending on the chosen parameterization \cite{gazeau2009, Bengtsson2006, radcliffe71}. In general, a spin coherent state refers to a rotation of the eigenstate of $J^z$ with a maximum eigenvalue of $m=j$
\begin{equation}
    \ket{\psi_\mathrm{SC}}=R\ket{j,j}.
\end{equation}
$R$ is a unitary rotation operator. This implicit definition can be made more explicit by realizing that $\ket{\psi_\mathrm{SC}}$ is an eigenstate of the rotation of $J^z$
\begin{equation}
    RJ^zR^\dagger\ket{\psi_\mathrm{SC}}=RJ^z\ket{j,j}=j\ket{\psi_\mathrm{SC}}.
\end{equation}
Since $R$ is a rotation there exists a unit vector $\vec{n}\in S^2$ such that $RJ^zR^\dagger=\vec{n}\cdot\vec{J}$. Thus, up to a phase, any spin coherent state can be characterized by a point on the sphere $\ket{\psi_{SC}}\equiv\ket{\vec{n}}$, with
\begin{equation}
    \vec{n}\cdot\vec{J}\ket{\vec{n}}=j\ket{\vec{n}}.
\end{equation}
For a unique definition we have to assume some convention, or gauge, in order to specify the phase of these states. This can be achieved for example in the Bargmann parameterization of spin coherent states, related to the stereographic projection of the sphere onto the complex plane by $z=\tan\frac{\theta}{2}\eul^{\ii\phi}$, where $\theta$ and $\phi$ are the polar angels pointing towards $\vec{n}$. In this parameterization a (unnormalized) spin coherent state can be explicitly written as \cite{radcliffe71,gazeau2009}
\begin{equation}\label{eq:spinbargmann}
    \ket{z}=\eul^{zJ^-}\ket{j,j} ,\qquad j\vec{n}=\frac{\braket{z|\vec{J}|z}}{\braket{z|z}} .
\end{equation}
This represents an analogy to bosonic Bargmann coherent states \cite{Bargmann62}. The above definition already specifies a phase of the states. However, the spin coherent state $\ket{j,-j}$, corresponding to $\vec{n}=-\vec{e}_z$ which is orthogonal to $\ket{j,j}$, is not included, as this would correspond to $|z|=\infty$ and the phase at this point is ill-defined. In fact, due to the nontrivial topology of the space of pure spin coherent states, it is impossible to find a smooth global phase gauge \cite{Mosseri2001Nov}. While the overlap of two spin coherent states depends on the phases, the absolute value of this overlap is gauge independent and given as
\begin{equation}
   |\braket{\vec{n}|\vec{n}'}|^2=\left(\frac{1+\vec{n}\cdot\vec{n}'}{2}\right)^{2j} ,
\end{equation}
where $\ket{\vec{n}}$ and $\ket{\vec{n}'}$ are assumed to be normalized. 
While any set of $2j+1$ distinct coherent states already spans the full spin Hilbert space \cite{Chryssomalakos2018}, the standard overcompleteness relation from the general group theoretical construction of coherent states is most commonly used \cite{Brif99}
\begin{equation}\label{eq:SCoherent_completeness}
    \id =\int_{S^2}\diff\Omega \ketbra{\vec{n}}{\vec{n}} ,
\end{equation}
with the proper uniform measure on $S^2$
\begin{equation}\label{eq:dOmega}
    \diff\Omega=\frac{2j+1}{4\pi}\diff^3{n}\,\delta(|\vec{n}|-1)=\frac{2j+1}{4\pi}\diff\phi\,\diff\theta\,\sin{\theta}.
\end{equation}
Similar to NMQSD for bosonic baths, this identity relation proves to be an elegant tool for developing a stochastic unraveling of open system dynamics from the microscopic model of system and bath.

\section{spin NMQSD Theory}\label{sec:stochastic_theory}

We consider model \eqref{eq:spinbathH} and wish to establish a stochastic theory similar to NMQSD based on spin coherent states and in particular on the completeness relation \eqref{eq:SCoherent_completeness}. We will first consider a zero temperature bath and elaborate on the finite temperature case later. Our starting point is the partial trace with respect to the environment expressed in terms of coherent states:
\begin{equation}
    \tr_E\ketbra{\Psi(t)}{\Psi(t)}=\int\left(\prod_{\lambda=1}^N \diff\Omega_\lambda\right)\braket{\boldsymbol{\vec{n}}|\Psi(t)}\braket{\Psi(t)|\boldsymbol{\vec{n}}}
\end{equation}
The bold vector $\boldsymbol{\vec{n}}=(\vec{n}_\lambda)$ is a shorthand notation for the coherent state labels of all bath spins, and $\diff \Omega_\lambda$ is as in \eqref{eq:dOmega}. The system Hilbert space vectors $\braket{\boldsymbol{\vec{n}}|\Psi(t)}$ can be used as a definition for pure state trajectories. However, they are not normalized which is a severe issue for Monte Carlo sampling of trajectories. After a short time, a large majority of trajectories may have a vanishing contribution to the average whereas single, rare trajectories contribute with a significant weight. To find an average over normalized states we divide by the norm to find
\begin{equation}\label{eq:SCunrav_linear}
        \tr_E\ketbra{\Psi(t)}{\Psi(t)}=\quad\int\left(\prod_{\lambda=1}^N \diff\Omega_\lambda\right)Q(t,\boldsymbol{\vec{n}})\frac{\braket{\boldsymbol{\vec{n}}|\Psi(t)}\braket{\Psi(t)|\boldsymbol{\vec{n}}}}{\braket{\Psi(t)|\boldsymbol{\vec{n}}}\braket{\boldsymbol{\vec{n}}|\Psi(t)}},
\end{equation}
where we defined the function $Q(t,\boldsymbol{\vec{n}})=\braket{\Psi(t)|\boldsymbol{\vec{n}}}\braket{\boldsymbol{\vec{n}}|\Psi(t)}$. In fact, this squared norm of the system Hilbert space vectors $\braket{\boldsymbol{\vec{n}}|\Psi(t)}$ is the spin $Q$-function (phase space distribution) of the reduced state of the environment.  To proceed, we can derive an evolution equation for this function. The derivation is provided explicitly in appendix \ref{app:Qfunc}. Due to the linear coupling and linear bath Hamiltonian, the $Q$-function satisfies a Liouville equation
\begin{equation}\label{eq:SpinQLangevin}
    \partial_tQ(t,\boldsymbol{\vec{n}})=-\sum_{\lambda=1}^N \vec{\nabla}_\lambda\cdot \vec{A}_\lambda(t,\vec{n}_\lambda) Q(t,\boldsymbol{\vec{n}}) ,
\end{equation}
with $\nabla_\lambda^a=\frac{\partial}{\partial n_\lambda^a}$ and the \emphquote{drift} term
\begin{equation}
    \vec{A}_\lambda(t,\vec{n}_\lambda)=g_\lambda O_\lambda(t)^T\braket{\vec{L}}(t)\times \vec{n}_\lambda.
\end{equation}
$\braket{\vec{L}}(t)={\braket{\Psi(t)|{\boldsymbol{\vec{n}}}}\vec{L}\braket{\boldsymbol{\vec{n}}|\Psi(t)}}/{\braket{\Psi(t)|\boldsymbol{\vec{n}}} \braket{\boldsymbol{\vec{n}}|\Psi(t)}}$ is the expectation value of the coupling operators with respect to the system state $\braket{\boldsymbol{\vec{n}}|\Psi(t)}$. As should be, the characteristic curves of this first order partial differential equation remain on the sphere. The solution is given by transport of the initial condition along the characteristic curves
\begin{equation}\label{eq:characteristic_eq}
    \dot{\vec{n}}_\lambda'(t)=\vec{A}_\lambda(t,\vec{n}_\lambda'(t)) ,\qquad \vec{n}_\lambda'(0)=\vec{n}_\lambda.
\end{equation}
One can identify the classical equation of motion for the bath spins due to coupling to the system state $\braket{\vec{\boldsymbol{n}}|\Psi(t)}$. It is surprising that these classical equations appear in this exact theory even though the bath evolution is not Gaussian. Using the method of characteristics we can explicitly write the $Q$-function as
\begin{equation}\label{eq:q_sol}
    Q(t,\boldsymbol{\vec{n}})=\int\left(\prod_{\lambda=1}^N \diff\Omega_\lambda''\,\delta(\vec{n}_\lambda''-\vec{n}_\lambda'(t))\right)Q(0,\boldsymbol{\vec{n}''}) .
\end{equation}
When inserted in \eqref{eq:SCunrav_linear}, this yields an average of normalized trajectories with time independent weights
\begin{equation}\label{eq:SCunrav_nonlinear}
        \tr_E\ketbra{\Psi(t)}{\Psi(t)}=
        \int\diff\mu(\boldsymbol{\vec{n}})\frac{\braket{\boldsymbol{\vec{n}}'(t)|\Psi(t)}\braket{\Psi(t)|\boldsymbol{\vec{n}}'(t)}}{\braket{\Psi(t)|\boldsymbol{\vec{n}}'(t)}\braket{\boldsymbol{\vec{n}}'(t)|\Psi(t)}}.
\end{equation}
In equation \eqref{eq:SCunrav_linear} each pure state trajectory belongs to a fixed set of coherent state labels and has a time dependent weight. Through the transformation \eqref{eq:q_sol} resulting in equation \eqref{eq:SCunrav_nonlinear} the coherent state labels now follow the evolution of the characteristic curves of the environment $Q$-function, but the weight of each trajectory becomes time independent. 
For a zero temperature bath initial state the coherent state labels $\vec{n}_\lambda$ are distributed according to the non-Gaussian measure
\begin{equation}\label{eq:SCmeasure}
        \diff\mu(\boldsymbol{\vec{n}})=\left(\prod_{\lambda=1}^N\diff\Omega_\lambda\right)Q(0,\boldsymbol{\vec{n}})=\prod_{\lambda=1}^N\diff\Omega_\lambda\left(\frac{1+n_\lambda^z}{2}\right)^{2j}.
\end{equation}
Defining an average $\expval{...}$ as the integration with respect to $\diff\mu(\boldsymbol{\vec{n}})$ we can obtain the reduced state of the system via
\begin{equation}
    \rho_S(t)=\expval{\frac{\braket{\boldsymbol{\vec{n}}'(t)|\Psi(t)}\braket{\Psi(t)|\boldsymbol{\vec{n}}'(t)}}{\braket{\Psi(t)|\boldsymbol{\vec{n}}'(t)}\braket{\boldsymbol{\vec{n}}'(t)|\Psi(t)}}} ,
\end{equation}
as in Eq.~\eqref{eq:NMQSD_avg}. Remarkably, this allows us to determine the reduced state with Monte Carlo sampling of normalized pure states (\emphquote{stochastic states}) in the system Hilbert space, because every sample contributes to the mixed state evolution with the same weight. 

It remains to find an evolution equation for the NMQSD trajectories $\ket{\psi(t,\boldsymbol{\vec{n}})}=\braket{\boldsymbol{\vec{n}'}(t)|\Psi(t)}$. For this a projection formalism in terms of a Feshbach partitioning can be used. This idea was introduced in Ref.~\cite{Link17}. It is based on finding a projection operator $P(t)$ in the full Hilbert space such that it extracts a single stochastic state and maps the initial state to itself. In our case the requirements are explicitly
\begin{equation}\label{eq:p_requirements}
    P(t)\ket{\Psi(t)}\propto \ket{\psi(t,\boldsymbol{\vec{n}})},\,\qquad P(0)\ket{\vac}=\ket{\vac}.
\end{equation}
As shown in Ref.~\cite{Link17}, applying a standard Feshbach method with such a projector gives a closed evolution equation for the stochastic state. When defining this projector in the spin case, we encounter the problem that the spin coherent state $\ket{\boldsymbol{\vec{n}'}(t)}$ is not smoothly defined for all values of $\vec{n}'_\lambda(t)$, because no smooth global phase gauge exists. To circumvent this issue, we define a time-dependent rotation operator as
\begin{equation}\label{eq:ROp}
    \partial_tR(t)=-\ii \braket{\vec{L}}(t)\cdot\sum_{\lambda=1}^N g_\lambda O_\lambda(t)\vec{J}_\lambda R(t) ,
\end{equation}
with initial condition $R(0)=\id$. Since $R(t)$ is a rotation it acts on the spin operators as $R^\dagger(t)\vec{J}_\lambda R(t)=S_\lambda(t)\vec{J}_\lambda$, where $S_\lambda(t)\in \mathrm{SO}(3)$ are orthogonal matrices. Note that this rotation propagates the coherent state $\ket{\boldsymbol{\vec{n}}}$ to the shifted state $\ket{\boldsymbol{\vec{n}'}(t)}=R(t)\ket{\boldsymbol{\vec{n}}}$ or equivalently $S_\lambda(t)\vec{n}_\lambda=\vec{n}_\lambda'(t)$. Because the dynamics of the phase of the shifted state is smoothly specified by the rotation operator $R(t)$, the phase singularity issue is circumvented. We note in passing that by writing $\ket{\boldsymbol{\vec{n}'}(t)}$ we abuse the notation because the phase of this state is not determined by the orientations $\vec{n}_\lambda$ alone. With this construction we can make the following choice for a time dependent projector
\begin{equation}\label{eq:projector}
    P(t)=\frac{R(t)\ketbra{\vac}{\boldsymbol{\vec{n}}}R^\dagger(t)}{\braket{\boldsymbol{\vec{n}}|\vac}}.
\end{equation}
This operator satisfies all desired properties required in the stochastic Feshbach theory. $P(t)$ is well defined by construction because spin coherent states orthogonal to the initial state $\ket{\vac}$ have zero probability in the average \eqref{eq:SCunrav_nonlinear}. It is a projector for all times $P(t)^2=P(t)$ and the relations \eqref{eq:p_requirements} hold. Using this operator, the Feshbach theory can be readily applied to find evolution equations for the stochastic state. The derivation is provided in appendix \ref{app:feshbach} and the resulting equation reads
\begin{equation}\label{eq:Feshbach}
\begin{split}
        \partial_t \ket{\psi(t)}=&-\ii H_S\ket{\psi(t)}-\ii \Delta\vec{L}(t)\cdot\sum_{\lambda=1}^N g_\lambda O_\lambda(t)S_\lambda(t)\frac{\braket{{\vec{n}_\lambda}|\vec{J}_\lambda|j,j}}{\braket{{\vec{n}_\lambda}|j,j}}\ket{\psi(t)}-\int_0^t\diff s\, K(t,s)\ket{\psi(s)},
\end{split}
\end{equation}
where we introduced the abbreviation $\Delta\vec{L}(t)=\vec{L}-\braket{\vec{L}}(t)$. For a better readability we have omitted to write the dependencies on $\bvec{\vec{n}}$ explicitly ($\ket{\psi(t)}\equiv\ket{\psi(t,\bvec{\vec{n}})}$). The exact form of the kernel operator $K(t,s)$ is given in the appendix. In this equation the non-Markovian nature of the bath is apparent via the convolution term which is not local in time. In general, the kernel operator is difficult to determine except within perturbation theory. In fact, while the projector approach is very transparent and elegant, the hierarchy expansion discussed in a later section may be more relevant in practice, as it can be directly employed to numerically solve for the stochastic state.

Finally, we want to show how to describe thermal initial states within the same framework. At finite temperature $k_B T=1/\beta$ the initial state of the bath in Hamiltonian \eqref{eq:spinbathH} is given by
\begin{equation}
 \rho_E(\beta)=\frac{1}{Z}\bigotimes_{\lambda=1}^N\eul^{-\beta \omega_\lambda (j-J_\lambda^z)} 
\end{equation}
and the partition function reads
\begin{equation}
    Z=\prod_\lambda \frac{1-\eul^{-\beta\omega_\lambda (2j+1)}}{1-\eul^{-\beta\omega_\lambda}}.
\end{equation}
Note that this termal state can also be expressed in terms of a distribution of coherent states by using the so-called $P$ representation
\begin{equation}
 \rho_E(\beta)=\int\left(\prod_{\lambda=1}^N\diff\Omega_\lambda\right)\,P_\beta(\boldsymbol{\vec{m}})\ket{\boldsymbol{\vec{m}}}\!\bra{\boldsymbol{\vec{m}}},
\end{equation}
where $\boldsymbol{\vec{m}}=(\vec{m}_\lambda)$ denotes a set of coherent state labels for the bath spins. The $P$-function can be explicitly computed and reads \\
\begin{equation}\label{eq:Pfunc}
\begin{split}
     &P_\beta(\boldsymbol{\vec{m}})=\frac{1}{Z}\prod_{\lambda=1}^N{\eul^{\beta \omega_\lambda}}\left(\frac{1+\eul^{\beta\omega_\lambda}}{2}+m_\lambda^z\frac{1-\eul^{\beta\omega_\lambda}}{2}\right)^{-2j-2}.
\end{split}
\end{equation}
Since this function is positive the thermal initial state can be realized by averaging over pure state dynamics corresponding to coherent initial states \cite{ Hartmann2017,Link17}. Note that coherent initial states are just rotations of the zero temperature bath state. A fixed rotation of the bath spins does not modify the linear form of the Hamiltonian, so that this can be captured by our theory. To be more explicit, let $\expvalbeta{...}$ denote the average of coherent state labels $\boldsymbol{\vec{m}}$ with respect to the thermal $P$-function, so that $\expvalbeta{\ketbra{\boldsymbol{\vec{m}}}{\boldsymbol{\vec{m}}}}=\rho_E(\beta)$. Then, the full system and bath state $\rho_{SE}$ can be obtained in the following way as an average over dynamics with coherent initial states
\begin{equation}
\rho_{SE}(t)=U(t)\ket{\psi_0}\rho_E(\beta)\bra{\psi_0} U^\dagger(t)=\expvalbeta{U(t)\ket{\psi_0}\ket{\boldsymbol{\vec{m}}}\bra{\boldsymbol{\vec{m}}}\bra{\psi_0} {}U^\dagger(t)}.
\end{equation}
$U(t)$ denotes the time evolution operator for the full system-and-bath state. By moving to a rotated frame, the coherent initial conditions can be set to the zero temperature bath state. In particular, let $R(\boldsymbol{\vec{m}})$ be a rotation operator which satisfies $R(\boldsymbol{\vec{m}})\ket{\boldsymbol{0}}=\ket{\boldsymbol{\vec{m}}}$. Then the Hamiltonian in the rotated frame is transformed as $H(t,\boldsymbol{\vec{m}})=R^\dagger(\boldsymbol{\vec{m}})H(t)R(\boldsymbol{\vec{m}})$ and reads
\begin{equation}\label{eq:HamiltonianNonzeroT}
    H(t,\boldsymbol{\vec{m}})=H_S+\sum_{\lambda=1}^N g_\lambda \vec{L}\cdot O_\lambda(t)M(\vec{m}_\lambda)\vec{J}_\lambda,
\end{equation}
where $M(\vec{m}_\lambda)\in \mathrm{SO}(3)$ satisfies $M(\vec{m}_\lambda)\vec{e}_z= \vec{m}_\lambda$. In summary, to realize a thermal initial condition of the bath one can propagate the state with zero temperature initial condition for a set of rotated Hamiltonians \eqref{eq:HamiltonianNonzeroT}, where the rotation is drawn randomly according to the $P$-function \eqref{eq:Pfunc}. In practice this can be achieved by Monte Carlo sampling of the labels $\vec{m}_\lambda$. Crucially, the form of Hamiltonian \eqref{eq:HamiltonianNonzeroT} is identical to \eqref{eq:hamiltonian}. Thus, all results from the zero temperature case can still be utilized. Since NMQSD is a stochastic method itself, such a stochastic realization of the thermal initial state poses no further complication \cite{Hartmann2017}.

\section{Pure Dephasing}\label{sec:pure_dephasing}
\begin{figure}[t]\centering
\includegraphics[height=\figheight]{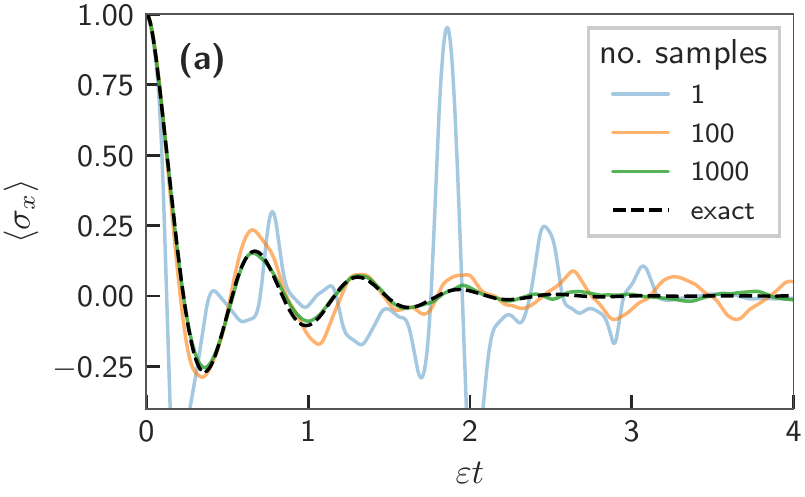}\qquad \includegraphics[height=\figheight]{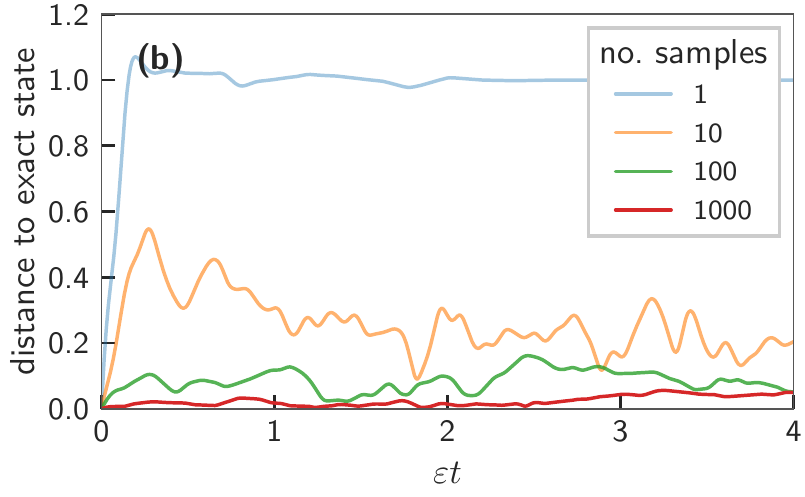}
\caption{Pure dephasing dynamics (Hamiltonian \eqref{eq:tls_dephasing_H}, $j=1/2$) in a spin bath with parameters $\omega_\lambda$, $g_\lambda$ from a discretized Ohmic SD (Eq.~\eqref{eq:ohmic_SD}, $s=1$, $\alpha=8$, $\omega_c=10\varepsilon$) with $N=100$ frequencies chosen linear from 0 to $60\varepsilon$ and finite temperature $\beta\varepsilon=0.5$. The initial state of the system is $\ket{\psi_0}=(\ket{0}+\ket{1})/\sqrt{2}$. (a) Estimated evolution of $\braket{\sigma_x}$ for different numbers of NMQSD trajectories in the average \eqref{eq:SCunrav_nonlinear}. (b) Convergence of spin NMQSD to the exact solution. The Hilbert-Schmidt distance between the exact reduced state and the ensemble average vanishes for a large number of trajectories.}\label{fig:spin_NMQSD_dephasing_ohmic_1}
\end{figure}

To demonstrate the validity of the spin NMQSD theory as well as the stochastic realization of finite temperature, we first consider a solvable model, namely pure dephasing in a spin bath, i.e.~Hamiltonian \eqref{eq:spinbathH} with $[L^x,H_S]=0$ and $L^y=L^z=0$ \cite{BhaktavatsalaRao2011Mar,Hsieh2018Jan}. In this model the evolution of the full system and bath state can easily be derived analytically, as shown in appendix \ref{app:pure_dephasing}. Therefore, we can use this as a test for the unraveling. For pure dephasing we do not need to solve the intricate equation of motion \eqref{eq:Feshbach} to find the stochastic states. Instead, we can simply extract these states directly from the exact system and bath state evolved with the transformed Hamiltonian \eqref{eq:HamiltonianNonzeroT}. This state has the form
\begin{equation}
    \ket{\Psi(t,\bvec{\vec{m}})}=\sum_n \eul^{-\ii \varepsilon_n t}\braket{\varepsilon_n|\psi_0}\ket{\varepsilon_n}\bigotimes_{\lambda=1}^N\ket{\psi_{\lambda,n}(t,\vec{m}_\lambda)}.
\end{equation}
The summation above is with respect to all system energies $H_S\ket{\varepsilon_n}=\varepsilon_n\ket{\varepsilon_n}$ and the states $\ket{\psi_{\lambda,n}(t,\vec{m}_\lambda)}$ are spin coherent states of the individual environment spins that depend on the stochastic thermal spin orientations $\vec{m}_\lambda$. The exact expression for a single NMQSD trajectory in the model can be obtained from projection of the full state onto the co-moving coherent state $\bra{\boldsymbol{\vec{n}}'(t)}$ (see appendix \ref{app:pure_dephasing}). The simplest nontrivial example that we can consider is a two level system coupled to a spin bath. The Hamiltonian for this model reads
\begin{equation}\label{eq:tls_dephasing_H}
    H=\varepsilon\sigma_z+\ketbra{1}{1}\sum_{\lambda=1}^N g_\lambda J_\lambda^x+\sum_{\lambda=1}^N \omega_\lambda (j-J_\lambda^z),
\end{equation}
where $\sigma_z=-\ketbra{0}{0}+\ketbra{1}{1}$. To specify the parameters of the bath we use a discretization of a continuous spectral density \cite{Bramberger2020Jan}. The spectral density of the bath is given by
\begin{equation}
    J(\omega)={\pi}\sum_{\lambda=1}^N g_\lambda^2 \delta(\omega-\omega_\lambda),
\end{equation}
which is the fourier transform of the bath correlation function $\alpha(t)$. As an example we choose couplings $g_\lambda$ and frequencies $\omega_\lambda$ according to a discretized Ohmic environment with spectral density
\begin{equation}\label{eq:ohmic_SD}
    J(\omega)=\frac{\pi}{2}\alpha \omega\eul^{-\omega/\omega_c}\Theta(\omega) ,
\end{equation}
see Ref.~\cite{Hartmann2019Jun} for details on discretization of continuous spectral densities. We consider a finite temperature which is realized using the stochastic method explained previously. With our example we can confirm that the ensemble average of thermal spin NMQSD trajectories converges to the exact reduced state as expected, see Fig.~\ref{fig:spin_NMQSD_dephasing_ohmic_1}. For the parameters considered here, 1000 samples give good convergence. Single trajectories show a typical pinning behavior known for QSD of pure dephasing: Trajectories localize close to an eigenstate of the system Hamiltonian. For an initial state with equal amplitude in each eigenstate, the trajectories localize with equal probability on all eigenstates so that on average the populations are conserved but the coherences decay \cite{Strunz99}.

\section{Hierarchy Expansion}\label{sec:hierarchy_expansion}
With a hierarchical expansion the equation of motion for the stochastic state \eqref{eq:Feshbach} can be systematically solved. The idea is similar to the hierarchy of pure states (HOPS) approach in Gaussian NMQSD \cite{suess14,suess15,Hartmann2017,Zhang2018Apr,Zhang2016Nov} or the related hierarchical equations of motions (HEOM) method \cite{Tanimura06,Tanimura14,Tang2015Dec,Tanimura2020Jul}. We would like to emphasize from the beginning that the hierarchy expansion for a spin bath could potentially demand considerably more resources in comparison to Gaussian baths. This is primarily because individual spins cannot be combined to form a single, larger spin (unlike bosonic modes).

To derive the hierarchy it is useful to switch to the Bargmann parameterization of spin coherent states (Eq.~\eqref{eq:spinbargmann}). For Bargmann states the following relations are easy to prove:
\begin{equation} \label{eq:BARGrelations}
\begin{split}
  &J_\lambda^x\ket{\vz}=j z_\lambda\ket{\vz}+\frac{1}{2}(1-z_\lambda^2)\partial_{z_\lambda}\ket{\vz}\\
  &J_\lambda^y\ket{\vz}=-\ii j z_\lambda\ket{\vz}+\frac{1}{2\ii}(-1-z_\lambda^2)\partial_{z_\lambda}\ket{\vz}\\
  &J_\lambda^z\ket{\vz}=j\ket{\vz}-z_\lambda\partial_{z_\lambda}\ket{\vz}
\end{split}
\end{equation} 
These allow us to write the action of spin operators in a differential form. We can redefine the norm of the stochastic states in order to express them with the Bargmann states $\braket{\vz|R^\dagger(t)|\Psi(t)}\equiv \ket{\psi(t,\boldsymbol{\vec{n}})}$. Taking the time derivative and making use of the relations \eqref{eq:BARGrelations} yields a differential equation for the stochastic state including derivatives with respect to the state label 
\begin{equation}\label{eq:hierarchy_0}
\begin{split}
         \partial_t\ket{\psi(t)}=&-\ii H_S\ket{\psi}-\ii\,\Delta\vec{L}(t)\cdot \sum_{\lambda=1}^N g_\lambda jO_\lambda(t) S_\lambda(t) \begin{pmatrix}
                                                        z_\lambda^*\\
                                                        \ii z_\lambda^*\\
                                                        1
                                                       \end{pmatrix}\ket{\psi(t)}\\&-\ii\,\Delta\vec{L}(t)\cdot\sum_{\lambda=1}^N g_\lambda O_\lambda(t) S_\lambda(t)\begin{pmatrix}
                                         \frac{1}{2}(1-(z_\lambda^*)^2)\\
                                         -\frac{1}{2\ii}(-1-(z_\lambda^*)^2)\\
                                         -z^*_\lambda
                                        \end{pmatrix}\partial_{z^*_\lambda}\braket{\vz|R^\dagger(t)|\Psi(t)}
\end{split}
\end{equation}
It turns out that the first two terms are exactly the first two terms of \eqref{eq:Feshbach}, thus the last term is the intricate memory term. Deriving an evolution equation for the derivative terms $\partial_{z^*_\lambda}\braket{\vz|R^\dagger|\Psi}$ one finds that higher order derivatives with respect to the Bargmann labels occur. We label these \emphquote{auxiliary states} with a multi-index $\bvec{k}\in \mathbb{N}_0^N$ as
\begin{equation}
    \ket{\psi^{{\bvec{k}}}(t,\boldsymbol{\vec{n}})}=\left(\prod_{\lambda=1}^N \partial_{z_\lambda^*}^{k_\lambda}\right)\braket{\vz|R^\dagger(t)|\Psi(t)}.
\end{equation}
To find the evolution equation for all auxiliary states we can simply take the corresponding $z^*_\lambda$ derivatives of \eqref{eq:hierarchy_0}. Then, the following commutations can be used to move the derivatives to the right hand side
\begin{equation}
    \begin{split}
         &[\partial_{z_\lambda^*}^k,z_\lambda^*]=k\partial_{z_\lambda^*}^{k-1} , \\
         &[\partial_{z_\lambda^*}^k,(z_\lambda^*)^2]=2kz_\lambda^*\partial_{z_\lambda^*}^{k-2}+k(k-1)\partial_{z_\lambda^*}^{k-1}.
    \end{split}
\end{equation}
One obtains a hierarchy of equations of motion where states of \emphquote{order} $|\bvec{k}|=\sum_{\lambda=1}^N k_\lambda$ couple only to states one order above or below. The equations of motion read explicitly:
\begin{equation}\label{eq:hir}
\begin{split}
  \partial_t\ket{\psi^{\bvec{k}}(t)}=&-\ii H_S\ket{\psi^{\bvec{k}}(t)}-\ii\,\Delta\vec{L}(t)\cdot\sum_{\lambda=1}^N g_\lambda(j-k_\lambda) O_\lambda(t) S_\lambda(t)\begin{pmatrix}
                                                        z_\lambda^*\\
                                                        \ii z_\lambda^*\\
                                                        1
                                                       \end{pmatrix}\ket{\psi^{\bvec{k}}(t)}\\
  &-\ii\,\Delta\vec{L}(t)\cdot\sum_{\lambda=1}^N g_\lambda O_\lambda(t) S_\lambda(t)\begin{pmatrix}
                                         \frac{1}{2}(1-(z_\lambda^*)^2)\\
                                         -\frac{1}{2\ii}(-1-(z_\lambda^*)^2)\\
                                         -z_\lambda^*
                                        \end{pmatrix}\ket{\psi^{\bvec{k}+\bvec{e}_\lambda}(t)}
      \\&-\ii\,\Delta\vec{L}(t)\cdot\sum_{\lambda=1}^N k_\lambda\big(j-\frac{1}{2}(k_\lambda-1)\big) g_\lambda O_\lambda(t) S_\lambda(t)\begin{pmatrix}
                                                        1\\
                                                        \ii\\
                                                        0
                                                       \end{pmatrix}\ket{\psi^{\bvec{k}-\bvec{e}_\lambda}(t)}
\end{split}
\end{equation}
In this equation, $\bvec{e}_\lambda=(\delta_{\lambda\gamma})$ is a unit vector. The $\boldsymbol{k}=0$ state of this hierarchy is the stochastic state which we aim to determine. The full hierarchy is as large as the full Hilbert space of system and bath, i.e.~exponentially large. Therefore, the hierarchy must be truncated at a low order to be numerically feasible, as in the case of HOPS. Formally, the truncation of the hierarchy at finite order $|\boldsymbol{k}|$ corresponds to a perturbative expansion in $\sum_{\lambda=1}^N g_\lambda^2$ to the same order. Note, however, that the auxiliary states with $k_\lambda=2j+1$ do not couple to the lower states and thus they remain zero and the hierarchy truncation becomes exact after $k_\lambda=2j$. In contrast, a standard perturbative expansion would require infinite orders to recover the exact solution. In fact, it is known for Gaussian baths that truncated hierarchical expansions can be used even for ultra strong coupling problems where perturbation theory is inherently unjustified \cite{Xu2022Nov}. The simplest sensible truncation for nontrivial spin bath problems is the second order truncation $|\boldsymbol{k}_{max}|=2$. At second order the hierarchy has $1/2(N+2)(N+1)\propto \mathcal{O}(N^2)$ auxiliary states, which is easily feasible also for larger values of $N$. A first order truncation does not make sense, as this would be consistent with Redfield theory where the spin bath could be replaced by a bosonic bath in the first place. Furthermore, in case of an extremely large number of spins $N$, for which the second order truncation would become intractable, we do not expect deviations from a Gaussian approximation of the bath except for fine tuned parameters or extremely strong coupling.

\begin{figure}\centering
\includegraphics[height=\figheight]{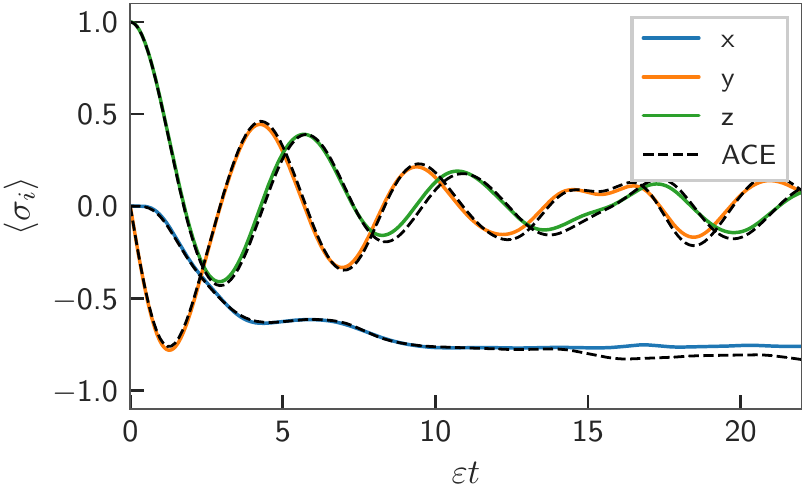}
\caption{Decay of a two level system in a spin bath (Hamiltonian \eqref{eq:exampleH}) computed with the hierarchy \eqref{eq:hir} in the second order truncation (average of 2000 samples). The bath consists of 60 spins with uniform randomly chosen frequencies and couplings $\omega_\lambda/\varepsilon\in [0,3]$, $g_\lambda^2/(N\varepsilon^2)\in [0,1.6]$. For comparison, the exact dynamics was computed with the ACE method \cite{Cygorek2021Jan} (dashed lines, convergence was reached with a discretization using 180 time steps and SVD compression with maximum bond dimension of 180). 
}\label{fig:spin_NMQSD_dephasing_hir}
\end{figure}

For a proof of concept we computed a nontrivial example problem which cannot be solved analytically. In particular, we considered a two level system  $H_S=\varepsilon \sigma_x/2$ coupled to a bath of two-level systems ($j=1/2$) with $\vec{L}=\sigma_z\vec{e}_x$, i.e.~the Hamiltonian
\begin{equation}\label{eq:exampleH}
    H=\frac{\varepsilon}{2} \sigma_x+\sigma_z\sum_\lambda g_\lambda J_\lambda^x +\sum_\lambda \omega_\lambda \left(\frac{1}{2}-J_\lambda^z\right).
\end{equation}
As is common for this model \cite{Hsieh2018Jan}, we choose couplings $g_\lambda$ and frequencies $\omega_\lambda$ randomly from a uniform distribution. In the spin-NMQSD simulation we used the second order hierarchy expansion. For comparison we computed the exact dynamics with the recently introduced ACE method (Automated Compression of Environments) \cite{Cygorek2021Jan}. This method is based on a representation of the discretized influence functional in terms of a tensor network, which has to be compressed iteratively with standard matrix product state compression techniques in order to ensure a feasible evaluation of the system dynamics. For the strong system-bath coupling that we choose in our example, ACE requires a large numerical effort reflected in a large bond dimension that is needed for a sufficiently accurate tensor network representation of the influence functional. This is an indication of strong coupling and a nontrivial structure of the environment. As can be seen in Fig.~\ref{fig:spin_NMQSD_dephasing_hir}, the approximate NMQSD solution from the second order hierarchy truncation is in good agreement with the ACE result. To improve the accuracy of the NMQSD calculation, a deeper hierarchy has to be considered. This would be feasible upon implementing more advanced flexible truncation schemes, as done for HOPS in Refs.~\cite{Zhang2018Apr, Varvelo2021}, or using matrix product state evolution methods as in Refs.~\cite{Flannigan2021Aug,Xu2022Nov,Gao2022Mar}. Compared to ACE, the hierarchy method has a more favourable scaling with respect to the size of the system Hilbert space, since it is based on wave function propagation instead of density matrices. However, a detailed analysis of the numerical performance of the method is beyond the scope of this paper and will be addressed in future works.

\section{Conclusions}\label{sec:conclusions}
In this article, we have presented a generalization of the non-Markovian quantum state diffusion method for arbitrary quantum systems coupled to linear spin baths. This advancement now enables the treatment of certain intricate many-body problems such as central spin models within the NMQSD framework. Our main contribution is the derivation of a representation for the time-evolved reduced state of the system as the average of an ensemble of pure states. These pure states are dependent on specific spin orientations $\vec{n}_\lambda$ that are distributed according to a non-Gaussian measure and can be sampled using Monte Carlo methods. This approach allows for an approximate determination of the reduced state by averaging over a finite set of stochastic pure states.
The labels $\vec{n}_\lambda$ thus take the role of the colored Gaussian noise process in the Gaussian NMQSD theory \cite{Strunz1996,Diosi1997}. Crucially, the stochastic states are normalized so that proper convergence with respect to the number of sampled states is assured. This kind of importance sampling requires that the time evolution of the stochastic states is nonlinear. In fact, the stochastic dynamics includes a co-moving semi-classical evolution of the bath spins via Eq.~\eqref{eq:characteristic_eq}. In general, for non-Markovian dynamics, determining evolution equations is difficult. With a projection approach we can formally construct such an evolution equation for the stochastic state, but an intricate memory kernel remains to be determined. Alternatively, we propose a hierarchical scheme to numerically propagate the stochastic states. Although the full hierarchy does not cure the exponential complexity of the problem at hand, we demonstrate with a simple example that a low order truncation is numerically feasible and gives good agreement with the exact dynamics. If, for instance in the case of strong coupling, a larger hierarchy is required to obtain accurate results, one could employ matrix product state techniques in order to keep the state dimension under control. Recently, such strategies have been successfully applied within the hierarchical expansion of Gaussian NMQSD \cite{Gao2022Mar,Flannigan2021Aug}. As a future perspective, it would be interesting to benchmark whether the proposed hierarchical scheme competes with other methods such as ML-MCTDH \cite{Wang2003Jul} or ACE \cite{Cygorek2021Jan} from a numerical perspective. In our view, the conditions under which a spin bath can be adequately described within a Gaussian framework \cite{Yang2016Nov} have not been investigated thoroughly in the literature. With the new range of methods available for treating non-Gaussian baths, it should now be possible to have a rigorous and quantitative discussion of this question.

\begin{acknowledgements}
VL acknowledges support from the international Max Planck research school (IMPRS) of MPI-PKS Dresden. VL and WTS are grateful for inspiring discussions with Chang-Yu Hsieh, Jianshu Cao and Shouryya Ray, as well as valuable comments from an anonymous Referee. Access to the computer resources of the Finnish IT Center for Science (CSC) and the FGCI project (Finland) is acknowledged by KL.
\end{acknowledgements}

\appendix
\section{Equation of Motion for the Q Function}\label{app:Qfunc}
To derive the equation of motion we crucially use the rules \cite{Merkel2020Dec}
\begin{equation}
\begin{split}
     &J^a\ketbra{\vec{n}}{\vec{n}}=jn^a\ketbra{\vec{n}}{\vec{n}}+\frac{1}{2}(\delta^{ab}-\ii\varepsilon^{abc}n^c-n^an^b)\partial^b \ketbra{\vec{n}}{\vec{n}}\\
     &\ketbra{\vec{n}}{\vec{n}}J^a=jn^a\ketbra{\vec{n}}{\vec{n}}+\frac{1}{2}(\delta^{ab}+\ii\varepsilon^{abc}n^c-n^an^b)\partial^b \ketbra{\vec{n}}{\vec{n}}.
\end{split}
\end{equation}
The $Q$ function can be written as $Q(t,\vec{n})=\tr \ketbra{\boldsymbol{\vec{n}}}{\boldsymbol{\vec{n}}}\rho_{SE}(t)$, where $\rho_{SE}(t)=\ketbra{\Psi(t)}{\Psi(t)}$ is the full state of system and environment. The derivation of the evolution equation goes as follows:
\begin{equation}
\begin{split}
  &\partial_tQ(t,\boldsymbol{\vec{n}})=-\ii\,\tr \left(\ketbra{\boldsymbol{\vec{n}}}{\boldsymbol{\vec{n}}}[H(t),\rho_{SE}(t)]\right)\\
  &=\sum_{\lambda=1}^N g_\lambda\varepsilon^{abc}n_\lambda^c\partial_{\lambda}^b\,\tr \left(\ketbra{\boldsymbol{\vec{n}}}{\boldsymbol{\vec{n}}}(O_\lambda(t)^T\vec{L})^a\rho_{SE}(t)\right)\\
  &=\sum_{\lambda=1}^N g_\lambda\varepsilon^{abc}n_\lambda^c\partial_{\lambda} ^b\,(O_\lambda(t)^T\braket{\vec{L}}(t))^a\,Q(t,\boldsymbol{\vec{n}})\\
  &=-\sum_{\lambda=1}^N g_\lambda\vec{\nabla}_\lambda\cdot \left( O_\lambda(t)^T\braket{\vec{L}}(t)\times \vec{n}_\lambda\right)\,Q(t,\boldsymbol{\vec{n}})
\end{split}
\end{equation}
Above, we have set $\partial_{\lambda}^a=\frac{\partial}{\partial n_\lambda^a}$ and defined
\begin{equation}
    \braket{\vec{L}}(t)=\frac{\tr \ketbra{\boldsymbol{\vec{n}}}{\boldsymbol{\vec{n}}}\vec{L}\rho_{SE}(t)}{\tr \ketbra{\boldsymbol{\vec{n}}}{\boldsymbol{\vec{n}}}\rho_{SE}(t)}=\frac{\braket{\Psi(t)|\boldsymbol{\vec{n}}}\vec{L}\braket{\boldsymbol{\vec{n}}|\Psi(t)}}{\braket{\Psi(t)|\boldsymbol{\vec{n}}}\braket{\boldsymbol{\vec{n}}|\Psi(t)}}.
\end{equation}

\section{Projection Formalism}\label{app:feshbach}
The derivation of a closed evolution equation for the stochastic state $\ket{\psi(t,\boldsymbol{\vec{n}})}$ is based on the projector \eqref{eq:projector}. The corresponding orthogonal projector is $Q(t)=\id-P(t)$. In the following we omit writing the dependence on the labels $\boldsymbol{\vec{n}}$. Taking the time derivative of the stochastic state and inserting an identity $\id=P(t)+Q(t)$ yields
\begin{equation}\label{eq:feshbachPPsi}
\begin{split}
        &\partial_t\ket{\psi(t)}=\braket{\boldsymbol{\vec{n}}|\left(\dot{R}^\dagger(t)-\ii R^\dagger(t) H(t)\right)P(t)|\Psi(t)}
        +\braket{\boldsymbol{\vec{n}}|\left(\dot{R}^\dagger(t)-\ii R^\dagger(t) H(t)\right)Q(t)|\Psi(t)}.
\end{split}
\end{equation}
The first term is easy to evaluate because $P\ket{\Psi}$ is proportional to the stochastic state itself. In order to find an expression for $Q\ket{\Psi}$ we can use the standard Feshbach technique and formally integrate the evolution equation 
\begin{equation}
\begin{split}
        &\partial_tQ(t)\ket{\Psi(t)}=\left(\dot{Q}(t)-\ii Q(t)H(t)\right)\ket{\Psi(t)}=\left(\dot{Q}(t)-\ii Q(t)H(t)\right)(Q(t)+P(t))\ket{\Psi(t)},
\end{split}
\end{equation}
which yields
\begin{equation}\label{eq:feshbachQPsi}
    \begin{split}
        &Q(t)\ket{\Psi(t)}=W(t,0)Q(0)\ket{\Psi(0)}+\int_0^t\diff s W(t,s)\left(\dot{Q}(s)
        -\ii Q(s)H(s)\right)P(s)\ket{\Psi(s)} ,
\end{split}
\end{equation}
where we introduced the time evolution operator $W(t,s)$ which is the solution to $W(s,s)=\id$ and
\begin{equation}
    \partial_t W(t,s)=\left(\dot{Q}(t)-\ii Q(t)H(t)\right)W(t,s).
\end{equation}
Note that the first term in \eqref{eq:feshbachQPsi} drops out because, by construction, the initial state lies in the subspace spanned by $P(0)$, so that $Q(0)\ket{\boldsymbol{0}}=0$. Inserting this formal solution into \eqref{eq:feshbachPPsi} and inserting the time derivatives of $R(t)$ from \eqref{eq:ROp} gives \eqref{eq:Feshbach}, where the kernel operator $K(t,s)$ reads
\begin{equation}
\begin{split}
         K(t,s)=&-\frac{1}{\braket{\boldsymbol{\vec{n}}|\boldsymbol{{0}}}}
        \braket{\boldsymbol{\vec{n}}|\left(\dot{R}^\dagger(t)-\ii R^\dagger(t) H(t)\right)\cdot W(t,s)\left(\dot{Q}(s)-\ii Q(s)H(s)\right)R(s)|{\boldsymbol{0}}}.
\end{split}
\end{equation}
A perturbation expansion of this operator is obtained upon inserting a perturbation expansion for $W(t,s)$. The lowest nonvanishing order is obtained from $W(t,s)\approx \id$.

\section{Pure Dephasing}\label{app:pure_dephasing}

We want to consider finite temperature of the bath, which can be realized stochastically as explained in the main text. Then, the pure dephasing Hamiltonian with a spin bath in the rotated interaction picture reads
\begin{equation}
    H(t,\boldsymbol{\vec{m}})=H_S+L\sum_{\lambda=1}^N g_\lambda \vec{e}_x\cdot O_\lambda(t)M(\vec{m}_\lambda)\vec{J}_\lambda .
\end{equation}
$H_S$ and $L$ commute so that the energy eigenstates $H_S\ket{\varepsilon_n}=\varepsilon_n\ket{\varepsilon_n}$ are also eigenstates of the coupling operator $L\ket{\varepsilon_n}=l_n\ket{\varepsilon_n}$. Projecting the Schrödinger equation onto an energy eigenstate then gives a closed equation in the bath Hilbert space\\
\begin{equation}
\begin{split}
    \ii\partial_t\braket{\varepsilon_n|\Psi(t,\boldsymbol{\vec{m}})}=&\varepsilon_n\braket{\varepsilon_n|\Psi(t,\boldsymbol{\vec{m}})}+l_n\sum_{\lambda=1}^N g_\lambda \vec{e}_x\cdot O_\lambda(t)M(\vec{m}_\lambda)\vec{J}_\lambda\braket{\varepsilon_n|\Psi(t,\boldsymbol{\vec{m}})}
\end{split}
\end{equation}
This corresponds to non-interacting spins so that the equation can be easily solved with a product ansatz
\begin{equation}
    \braket{\varepsilon_n|\Psi(t,\boldsymbol{\vec{m}})}= c_n\eul^{-\ii\varepsilon_nt} \bigotimes_{\lambda=1}^N \ket{\psi_{\lambda,n}(t,\vec{m}_\lambda)}.
\end{equation}
The evolution equation for the individual bath spin states conditioned on the system energy $\varepsilon_n$ are linear
\begin{equation}
    \ii\partial_t\ket{\psi_{\lambda,n}(t,\vec{m}_\lambda)}=g_\lambda l_n \vec{e}_x\cdot O_\lambda(t)M(\vec{m}_\lambda)\vec{J}_\lambda\ket{\psi_{\lambda,n}(t,\vec{m}_\lambda)}
\end{equation}
and, thus, spin coherent states are solutions. 
To find the stochastic state from the full solution we project the system and bath state onto spin coherent states
\begin{equation}
    \braket{\boldsymbol{\vec{n}}'(t)|\Psi(t,\boldsymbol{\vec{m}})}=\sum_n c_n\eul^{-\ii\varepsilon_n t}\ket{\varepsilon_n}\sum_{\lambda=1}^N \braket{\vec{n}_\lambda'(t)|\psi_{\lambda,n}(t,\vec{m}_\lambda)} ,
\end{equation}
where the time dependent labels $\vec{n}_\lambda '(t)$ are obtained by integrating 
\begin{equation}
    \dot{\vec{n}}_\lambda'(t)=
    \braket{L}(t)g_\lambda (M(\vec{m}_\lambda)^TO_\lambda(t)^T\vec{e}_x)
    \times \vec{n}_\lambda'(t).
\end{equation}
The initial conditions $\vec{n}_\lambda(0)$ and the thermal rotations $\vec{m}_\lambda$ can be drawn from the proper distributions \eqref{eq:SCmeasure} and \eqref{eq:Pfunc} using inverse transform sampling. 

For a comparison one can obtain the exact evolution at finite temperature by considering the evolution equation of the full system and bath state $\rho_{SE}(t)$ projected onto the energy eigenstates
\begin{equation}
    \ii\partial_t\braket{\varepsilon_n|\rho_{SE}(t)|\varepsilon_m}=\braket{\varepsilon_n|H|\varepsilon_n}\braket{\varepsilon_n|\rho_{SE}(t)|\varepsilon_m}-\braket{\varepsilon_n|\rho_{SE}(t)|\varepsilon_m}\braket{\varepsilon_m|H|\varepsilon_m}.
\end{equation}
The Hamiltonian here is the original full Hamiltonian \eqref{eq:tls_dephasing_H}. This equation can be solved by uncoupled spins using a product ansatz for $\braket{\varepsilon_n|\rho_{SE}(t)|\varepsilon_m}$. Thus, the computational effort is just linear in $N$ and the full dynamics can be easily computed because the spins are initially uncorrelated.

\bibliography{Bibliography.bib,bib.bib}

\end{document}